\documentclass[12pt,a4]{article}
\usepackage{hyperref,setspace}
\usepackage{amsthm,amsmath,latexsym,amssymb,amsfonts,amscd}
\usepackage{graphics}
\usepackage{tikz-cd}
\numberwithin{equation}{section}

\usepackage[numbers,sort&compress]{natbib}
\newcommand{\pl}{\partial}

\newcommand{\p}[1]{(\ref{#1})}
\newcommand{\be}{\begin{equation}}
\newcommand{\ee}{\end{equation}}
\newcommand{\bea}{\begin{eqnarray}}
\newcommand{\eea}{\end{eqnarray}}

\newcommand{\zb}{{\bar{z}}}

\newcommand{\pvec}{{\boldsymbol{p}}}
\newcommand{\qvec}{{\boldsymbol{q}}}
\newcommand{\kvec}{{\boldsymbol{k}}}
\newcommand{\pb}{{\bar{p}}}

\newcommand{\tr}{{\text{Tr}}}

\newcommand{\pfrac}[1]{{\frac{\pl}{\pl #1}}}

\newcommand{\PP}{{\mathbb{P}}}

\newcommand{\PPb}{{\overline{\mathbb{P}}}}

\newcommand{\veryshortarrow}[1][3pt]{\!\mathrel{%
   \vcenter{\hbox{\rule[-.2pt]{#1}{.4pt}}}%
   \mkern-4mu\hbox{\usefont{U}{lasy}{m}{n}\symbol{41}}}\!}

\begin{document}
%%%%%%%%%%%%%%%%%%%%%%%%%%%%%%%%%%%%%%%%%%%%%%%%%%%%%%%%%%
\begin{center}

\vskip 0.04\textheight

{\Large \bf   Supersymmetric Quantum Chiral Higher Spin Gravity 
 } \vskip .7in 
 {\Large 
Mirian Tsulaia\footnote{e-mail: {\tt  mirian.tsulaia@oist.jp }} and  
Dorin Weissman\footnote{e-mail: {\tt  dorin.weissman@oist.jp }} 
}
\vskip .4in {  \it  Okinawa Institute of Science and Technology, \\ 1919-1 Tancha, Onna-son, Okinawa 904-0495, Japan}\\
\vskip .8in

%\today

\vskip 0.4in

{\bf Abstract }

\end{center}
\begin{quotation}
We study quantum properties of supersymmetric
${\cal N}=1$ and ${\cal N}=4$ extensions of the
four dimensional bosonic
Chiral  Higher Spin Gravities (HiSGRAs).
We discuss the spectra,  the classical actions and
 define the  Feynman rules in 
${\cal N}=1$ and ${\cal N}=4$ superspaces in the light-front gauge.
Using these Feynman rules, we compute
tree and one-loop amplitudes for these systems.
A dimensional reduction to a system with ${\cal N}=2$
supersymmetry and  with massive higher spin fields is performed
and  quantum properties of this system are discussed.

\end{quotation}

\pagestyle{empty}
\newpage
\setcounter{page}{1}
\pagestyle{plain}
%\tableofcontents
\newpage

%%%%%%%%%%%%%%%%%%%%%%%%%%%%%%%%%%%%%%%%%%%%%%%%%%%%%%%%%%
\section{Introduction}
%%%%%%%%%%%%%%%%%%%%%%%%%%%%%%%%%%%%%%%%%%%%%%%%%%%%%%%%%%

Chiral Higher Spin Gravity (Chiral HiSGRA) has several unique
properties.
Chiral HiSGRA was originally introduced
in \cite{Ponomarev:2016lrm}, 
where it was shown that one can 
consistently ``truncate" a classical 
Hamiltonian of 
 \cite{Metsaev:1991mt}--\cite{Metsaev:1991nb},
in such a way that it stays purely
cubic without  any higher order corrections.
As a result, 
the action for the Chiral HiSGRA, which  is
 written in the light-front gauge,
 has  a simple local form,
 being in some sense 
a higher spin generalization  of the action for the self-dual Yang-Mills theory \cite{Chalmers:1996rq}\footnote{Further connections
between Chiral HiSGRA and self-dual theories were studied in
\cite{Ponomarev:2017nrr}.}.
The spectrum of the Chiral HiSGRA consists of an infinite tower of massless
fields with integer spins, each spin being present only once.
Among an infinite number of cubic interaction vertices present in the action, there are
familiar cubic vertices 
for lower spin fields, such  as three graviton chiral cubic vertex.
All these features
make  Chiral HiSGRA  the simplest higher spin extension of gravity and 
a candidate of being
 an essential building block of any
consistent interacting massless higher spin theory in four
${\cal D}=4$
dimensions.

The Quantum Chiral HiSGRA was then studied in 
\cite{Skvortsov:2018jea,Skvortsov:2020wtf,Skvortsov:2020gpn} where it was shown that despite it being naively non-renormalizable, 
the theory is consistent at the quantum level as well. The tree and loop amplitudes  
in the Chiral HiSGRA
 have a highly nontrivial structure, 
which nevertheless results in the $S$ matrix being $S=1$, in accordance
 with various no-go theorems
 \cite{Weinberg:1964ew,Coleman:1967ad} (see also \cite{Bekaert:2010hw,Sagnotti:2011jdy,Rahman:2015pzl}
 for reviews).
The tree  amplitudes vanish when putting all external momenta on shell,
as a result of nontrivial cancellations between individual Feynman diagrams. 
The loop
amplitudes vanish due to the presence of  an overall numerical factor $\nu_0$, which can be regularized to zero
according to the prescription given in \cite{Beccaria:2015vaa}.
An analogous situation, i.e. the presence of the overall factor $\nu_0$, happens for the other loop amplitudes as well. 
The peculiar form of the tree and loop amplitudes is due to a specific choice of the 
coupling constants in the cubic vertex, the so-called
``coupling constant conspiracy'', which 
allowed for the existence
of the classical Chiral HiSGRA in the first place.
 
 From ${\cal D}=4$ Chiral HiSGRA one can obtain some more
 consistent models with interacting higher spin fields, both on flat and on anti-de Sitter backgrounds.
 Namely, it is possible   to consider its  $AdS_4$ version
 and study its application to the $AdS_4/CFT_3$ correspondence 
 \cite{Vasiliev:1992av,
 Metsaev:2018xip,Skvortsov:2018uru, Sharapov:2022awp,
 Didenko:2022qga}.
 Another interesting model can be obtained by
 making a specific kind of dimensional reduction
 to a three dimensional ${\cal D}=3$ flat space
  \cite{Metsaev:2020gmb}--\cite{Skvortsov:2020pnk}.
 As it is common for three dimensional 
 systems with 
 higher spin fields (see for example
 \cite{Vasiliev:1995dn,
 Vasiliev:1999ba, Buchbinder:2015mta,  
 Kuzenko:2016bnv,Kuzenko:2016qwo, 
 Zinoviev:2021cmi, Khabarov:2022hbv} 
 and references therein),
 the model consists of massive higher spin fields,
 with an extra requirement that
   the  masses belong to 
 a particular lattice. The latter
 condition, along with the ``coupling constant conspiracy",
 ensures that the three dimensional action
  stays cubic, as it was in ${\cal D}=4$.
 
 Until now our discussion was concerned with the purely bosonic 
 Chiral HiSGRA. The aim of the present paper is to include
also fermionic fields into consideration. A natural way
to proceed is to start with four dimensional  supersymmetric
cubic vertices constructed in \cite{Bengtsson:1983pg,
Metsaev:2019dqt,Metsaev:2019aig}
\footnote{See 
\cite{Alkalaev:2002rq, Buchbinder:2017nuc, Hutomo:2017phh, 
Buchbinder:2018wzq,
Khabarov:2020deh, Gates:2019cnl, 
Buchbinder:2021qrg,
Buchbinder:2021igw,
Buchbinder:2021qkt}
for Lorentz  covariant  constructions for supersymmetric vertices in various dimensions.},
then keep their chiral part and choose a particular
form for the coupling constants, in order to make the Lagrangian purely cubic.
After that, one can develop a supersymmetric perturbation
theory for the Chiral Supersymmetric HiSGRAs in complete analogy
with its bosonic counterpart.

In the present paper we consider Chiral HiSGRAs with 
 ${\cal N}=1$
and ${\cal N}=4$ supersymmetries. Although it is possible
to construct cubic interaction vertices for massless higher spin
fields
also for the systems with higher number of supersymmetries
\cite{Bengtsson:1983pg,Metsaev:2019aig}
 (see also \cite{Devchand:1996gv}), 
 and to  develop  the perturbation theory for such systems, 
 here we restrict
 ourselves with more ``conventional"
 cases\footnote{Tree level scattering amplitudes for
 supersymmetric higher spin fields in the framework of higher spin-IKKT models were considered in
 \cite{Steinacker:2022jjv}. Tree level supersymmetric  scattering amplitudes with  massive higher spin exchanges  were recently discussed in \cite{KNBalasubramanian:2022sae}}.
 
At this point we would like to mention two more features of
the Chiral HiSGRA, which are heavily used throughout the paper.  
First, the Chiral HiSGRA allows for Chan-Paton-like gauging, with the same internal symmetry groups 
and the same  symmetry of the representations as in the Open String Theory \cite{Marcus:1982fr}. Taking a $U(N)$ ``colored" version
of the Chiral HiSGRA greatly simplifies the computations
since one can consider only color ordered amplitudes  \cite{Skvortsov:2018jea,Skvortsov:2020wtf,Skvortsov:2020gpn}.
Second, since the classical Chiral HiSGRA is formulated in the light-front
gauge,\footnote{A covariant version of the equations of motion
for the Chiral HiSGRA was recently obtained in \cite{Skvortsov:2022syz,
Sharapov:2022faa}.}
 for the computations of the amplitudes 
 we extend  the technique
developed in \cite{Chakrabarti:2005ny,Chakrabarti:2006mb}
for Yang-Mills theory to ${\cal N}=1$ and ${\cal N}=4$ 
light-front superspaces.

The paper is organised as follows.
In Section
\ref{sec:hisgra1}
we briefly describe  ${\cal D}=4$, ${\cal N}=1$ and ${\cal N}=4$
extensions
of the classical bosonic Chiral HiSGRA. We give the field content, the action
and the Feynman rules for these models. 
Section \ref{sec:trees} contains the results
of the computation of $n$-point\footnote{See \cite{Neiman:2022enh} 
for a recent developments for computation of $n$-point
diagrams for higher spin gravity in the framework
of $AdS_4/CFT_3$ correspondence.} tree
amplitudes, and Section \ref{sec:loops} presents the results for loop amplitudes.
In Section \ref{D3N2} we describe a dimensional reduction
to ${\cal D}=3$, ${\cal N}=2$ massive supersymmetric Chiral HiSGRA 
with central charges and a computation of tree
and one loop diagrams.
A brief summary of the derivation  of cubic vertices for 
${\cal D}=4$,
${\cal N}=1$ supersymmetric massless higher spin fields
\cite{Metsaev:2019dqt} 
and of a derivation
of the corresponding chiral model is given in the Appendix.

\section{ Classical \texorpdfstring{${\cal D}=4$}{D=4} Supersymmetric Chiral HiSGRA} \label{sec:hisgra1}
 
\subsection{\texorpdfstring{${\cal N}=1$}{N=1} Chiral HiSGRA}
\label{N1chiralhisgra}
The basic objects for the case of ${\cal N}=1$ supersymmetric
Chiral HiSGRA are superfields, which in general
can be either singlets or 
 belong to an algebra of matrices.
 The index $\lambda$
denotes a ''super helicity" i.e., a helicity of a superfield, 
and can be either integer or half--integer.
The  four--momentum $\pvec$
is split into longitudinal components 
$p^-\equiv \gamma$ and
$p^+\equiv \beta$
and to
a pair of mutually complex conjugated transverse components $p$
and $\bar p$. 
For the light-front superfield approach 
one introduces also Grassmann momenta $p_\theta$.

On the level of components  we have two sets of fields with integer and half-integer helicities,
 $\phi_\lambda(\pvec)$ and $\psi_\lambda(\pvec)$, combined into the superfields $\Theta_\lambda(\pvec, p_\theta)$
  as follows
\be \label{SFN1-1}
\Theta_{-s}(\pvec, p_\theta) =  \phi_{-s}(\pvec) + \frac{p_\theta}{\beta}\phi_{-s +\frac{1}{2}}(\pvec), 
\quad
\Theta_{s-\frac{1}{2}}(\pvec, p_\theta)  = -\phi_{s - \frac{1}{2}}(\pvec)
+ {p_\theta}\phi_{s}(\pvec), 
\ee
with $s=1,2,.., \infty$,
and 
\be \label{SFN1-2}
 \Theta_{s}(\pvec, p_\theta) =  \psi_{s}(\pvec) + \frac{p_\theta}{\beta}\psi_{s +\frac{1}{2}}(\pvec), 
\quad
 \Theta_{-s - \frac{1}{2}}(\pvec, p_\theta) = 
 -\psi_{-s - \frac{1}{2}}(\pvec)
+ {p_\theta}\psi_{-s}(\pvec), 
\ee
with $s=0, 1,2,.., \infty$.

The 
hermitian conjugation rules for the component
fields are defined as
\be
\phi^\dagger_\lambda(\pvec) = \phi_{-\lambda}(-\pvec), \quad \psi^\dagger_\lambda(\pvec) = \psi_{-\lambda}(-\pvec).
\ee 
The superfields    obey the   equal time Poisson brackets
\be\label{equaltime-theta}
    [\Theta_{\lambda}^{AB}(\pvec,p_\theta), \Theta_{\lambda^\prime}^{CD}(\pvec^\prime,p_\theta^\prime)]=-
    \delta_{\lambda +\lambda^\prime, -\frac{1}{2}}
    \frac{\delta^{3}(\pvec+\pvec^\prime)\,
    \delta(p_\theta+p_\theta^\prime)}
    {2\beta} \, \Pi^{AC,BD}_{(G)}\,,
 \ee
where $\Pi^{AC,BD}$ is a group theoretic factor whose explicit form depends on the choice of gauge group $G$.
The symmetry under the interchange of gauge indices also depends
on the gauge group.
Similarly
to how it has been done in  bosonic HiSGRA 
\cite{Metsaev:1991nb}
\cite{Skvortsov:2020wtf}, one can show that allowed gauge groups
include $U(N)$, $SO(N)$, and $USp(N)$, with\footnote{The indices of the $USp(N)$ group are raised and lowered in terms of antisymmetric matrices
$C_{AB}=-C_{BA}$, $C_{AB}C^{CB}=\delta^C_A$
as
$V^A=C^{AB}V_B$, $V^B C_{BA}=V_A$.}
\begin{align} \label{Piun}
  {\Pi_{(U(N))}}^{A,C}_{B,D}= &(-)^{\lambda+\frac{1}{2}\epsilon_\lambda}\delta^C_{  B}\delta^A_{ D}\, & 
 \end{align}
\begin{align} \label{Pison}
  \Pi^{AC,BD}_{(SO(N))}= &[\delta^{AC}\delta^{BD}+(-)^{\lambda + \frac{1}{2}\epsilon_\lambda }\delta^{AD}\delta^{BC}]\,, & 
  \Theta_{\lambda}^{AB}({\pvec}, p_\theta)&=(-)^{\lambda+ \frac{1}{2}\epsilon_\lambda}\Theta_{\lambda}^{BA}({\pvec}, p_\theta),
\end{align}
\begin{align} \label{Piuspn}
  \Pi^{AC,BD}_{(USp(N))}= &[C^{AC}C^{BD}-(-)^{\lambda + \frac{1}{2}\epsilon_\lambda }C^{AD}C^{BC}]\,, & 
  \Theta_{\lambda}^{AB}({\pvec}, p_\theta)&=-(-)^{\lambda+ \frac{1}{2}\epsilon_\lambda}\Theta_{\lambda}^{BA}({\pvec}, p_\theta).
\end{align}
 In the equations above we used the symbol
 \be \epsilon_\lambda \equiv \begin{cases} 0\,, \quad \lambda \in {\mathbb Z} \\ 1\,, \quad \lambda \in {\mathbb Z}+\frac12 \end{cases} \ee
With this notation, the Grassmann parity of a (super)field of (super)helicity $\lambda$ is $(-)^{\epsilon_\lambda}$.

The action for the  ${\cal N}=1 $ supersymmetric chiral HiSGRA is
\bea\label{eq:chiralaction}
S&=&- \frac{1}{2}\sum_{\lambda = -\infty}^\infty\int d^4 p \, d p_\theta \,\,  (-)^{\epsilon_{\lambda}}
(\pvec^2)\,
\mathrm{Tr}[\Theta_{\lambda-\frac{1}{2}}({\pvec, p_\theta})\,  \Theta_{-\lambda}({-\pvec, -p_\theta})] 
\\ \nonumber
&+&\sum_{\lambda_{1,2,3}}\int
\prod_{i=1}^3 d^4 p_i  \prod_{j=1}^3 d p_{\theta,j} \,
\delta^4(\pvec_1+\pvec_2+\pvec_3)
    \delta(p_{\theta_1}+p_{\theta_2}+p_{\theta_3})
C_{\lambda_1,\lambda_2,\lambda_3} V(\pvec_i,p_{\theta, i},\lambda_i),
\eea
where
\be\label{eq:generalvertex}
    V= \frac{\PPb^{\lambda_1+\lambda_2+\lambda_3+1}}
    {\beta_1^{\lambda_1 + \frac{1}{2}\epsilon_{\lambda_1}}
    \beta_2^{\lambda_2+ \frac{1}{2}\epsilon_{\lambda_2}}
    \beta_3^{\lambda_3+ \frac{1}{2}\epsilon_{\lambda_3}}}\tr
    [ \Theta_{\lambda_1}(\pvec_1, p_{\theta,1}) \,
 \Theta_{\lambda_2}(\pvec_2, p_{\theta,2}) \,
 \Theta_{\lambda_3}(\pvec_3, p_{\theta,3})].
   \ee
The trace in equations \p{eq:chiralaction} -- \p{eq:generalvertex}
is taken over the gauge group indices.
The sum of helicities in the expression \p{eq:generalvertex}
is constrained to be a non-negative integer, with other coupling constants vanishing.
The coupling constants $C_{\lambda_1,\lambda_2,\lambda_3}$ are chosen as
\begin{equation}\label{eq:magicalcoupling}
    C_{\lambda_1,\lambda_2,\lambda_3}=\frac{(-)^{\epsilon_{\lambda_2}}(l_p)^{\lambda_1+\lambda_2+\lambda_3}}{\Gamma(\lambda_1+\lambda_2+\lambda_3+1)}\, ,
\end{equation}
in order to make the action purely cubic 
\cite{Metsaev:2019dqt}
(see the Appendix for more details).
The action is chiral in the sense that the complex conjugated 
expression to the interaction term is absent.
Similarly to the bosonic chiral HiSGRA, the action contains chiral
parts of the known low spin cubic vertices, along
with an infinite number of vertices with higher spin fields.
In particular, the choice 
$\lambda_1=\lambda_2= \frac{1}{2}$, $\lambda_3 =-1$
gives the chiral part of the cubic vertices
for the ${\cal N}=1$ Super Yang-Mills.
Similarly,  the choice
$\lambda_1=\lambda_2= \frac{3}{2}$, $\lambda_3 =-2$
gives the
chiral part of the cubic vertices for the
${\cal N}=1$ Supergravity. The choice $\lambda_1=\lambda_2=\lambda_3=0$ 
corresponds to the chiral part of the cubic vertex
in the Wess-Zumino model \cite{Wess:1973kz} in the light
front gauge \cite{Mandelstam:1982cb} (the antichiral cubic coupling
being $\lambda_1=\lambda_2=\lambda_3=-\frac{1}{2}$).
This coupling
generates only interactions of a scalar with spin $\frac12$ fermions. The three scalar coupling is absent, as in the purely bosonic case.

As one can see from the field content given in the equations
\p{SFN1-1}-\p{SFN1-2}, 
 the ${\cal N}=1$ Chiral HiSGRA is not a straightforward supersymmetrization
 of the bosonic Chiral HiSGRA, since the spectrum
 of the former contains twice as many bosonic fields with nonzero helicities. The necessity for this ``doubling" of the spectrum
 is dictated by the fact that taking only one set (say, that given in equation \p{SFN1-2}),
 it would not have been possible to achieve the consistency
 at the level of quartic interactions and to obtain the chiral theory (see the discussion at the end of the Appendix).

\subsection{\texorpdfstring{${\cal N}=4$}{N=4} Chiral HiSGRA}\label{sec:hisgra4}
The discussion of the previous subsection can be appropriately
modified
to describe  ${\cal N}=4$ chiral HiSGRA.
In this case  one can 
consider higher spin 
superfields with only integer ``super helicities" $\lambda$.
These  superfields  have the form 
of the expansion in terms of  Grassmann momenta
$p_{\theta,{\hat i}}$, 
 with ${\hat i}=1,...,4$ \cite{Metsaev:2019aig}
\bea \label{superfieldN4}
\Theta_{\lambda}(\pvec, p_{\theta})& = &\beta  \phi_{\lambda-1}(\pvec) 
-  \phi_{\lambda -\frac{1}{2};{\hat i}}(\pvec)
p_{\theta, {\hat i}} + \frac{1}{2} 
\phi_{\lambda; {\hat i}{\hat j}}(\pvec) p_{\theta, {\hat i}}
p_{\theta, {\hat j}} - \\ \nonumber
&-&\frac{1}{3!}
\beta^{-1 } \phi_{\lambda +\frac{1}{2};{\hat i}}(\pvec)
\varepsilon^{{\hat i}{\hat j}{\hat k}{\hat l}} 
p_{\theta, {\hat j}}p_{\theta, {\hat k}}p_{\theta, {\hat l}}
+ \frac{1}{4!}\beta^{-1} \phi_{\lambda+1}(\pvec) 
\varepsilon^{{\hat i}{\hat j}{\hat k}{\hat l}} 
p_{\theta, {\hat i}}
p_{\theta, {\hat j}}p_{\theta, {\hat k}}p_{\theta, {\hat l}}
,
\eea
In particular, 
 for $\lambda=0$ the expression \p{superfieldN4} is
 the superfield for ${\cal N} =4$ Super Yang-Mills 
 in the light-front gauge   \cite{Mandelstam:1982cb}.
The component fields obey the hermitian conjugation properties
 \be
\phi^\dagger_{\lambda-1}(\pvec) = \phi_{-\lambda+1}(-\pvec), \quad \phi^\dagger_{\lambda-\frac{1}{2},{\hat i}}(\pvec) = \phi_{-\lambda-\frac{1}{2},{\hat i}}(-\pvec),
\quad
\phi^\dagger_{\lambda,{\hat i}{\hat j}}(\pvec) = 
\varepsilon^{{\hat i}{\hat j}{\hat k}{\hat l}} \phi_{-\lambda,{\hat k}{\hat l}}(-\pvec).
\ee 
For the equal-time Poisson brackets one has
 \be\label{equaltime-theta-4}
    [\Theta_{\lambda}(\pvec,p_\theta), \Theta_{\lambda^\prime}(\pvec^\prime,p_\theta^\prime)]=-
    \delta_{\lambda ,\lambda^\prime}
    \frac{\delta^{3}(\pvec+\pvec^\prime)\,
    \delta^4(p_\theta+p_\theta^\prime)}
    {2} \, \Pi^{AC,BD}_{(G)}
 \ee
 where the expressions for $\Pi^{AC,BD}_{(G)}$ are as in
 \p{Piun}-\p{Piuspn}.
 The action for the Chiral ${\cal N}=4$ HiSGRA is similar to the one for
 the Chiral ${\cal N}=1$ HiSGRA considered in the previous subsection,
  \bea\label{eq:chiralaction4}
S&=&-\frac{1}{2}\sum_{\lambda = -\infty }^\infty   \int d^4 p \, d^4 p_\theta \,\,
(\pvec^2)\,
\mathrm{Tr}[\Theta_{\lambda}({\pvec, p_{\theta, \hat i}})\,  \Theta_{-\lambda}({-\pvec, -p_{\theta. \hat i}})] 
\\ \nonumber
&+&\sum_{\lambda_{1,2,3}}\int
\prod_{i=1}^3 d^4 p_i  \prod_{j=1}^3 d^4 p_{\theta,j} \,
\delta^4(\pvec_1+\pvec_2+\pvec_3)
    \delta^4(p_{\theta_1}+p_{\theta_2}+p_{\theta_3})
C_{\lambda_1,\lambda_2,\lambda_3} V(\pvec_i, p_{\theta, i}, \lambda_i),
\eea
where the cubic vertex and the coupling constants are given in
\p{eq:generalvertex} and in
\p{eq:magicalcoupling}, respectively, with $\epsilon_{\lambda_i}=0$.
The action \p{eq:chiralaction4}  contains an infinite number of cubic vertices,
along with the chiral part of the  ${\cal N}=4$ Super Yang-Mills
cubic interactions \cite{Mandelstam:1982cb, Brink:1982pd}. The latter
can be obtained 
 from the action  \p{eq:chiralaction4} by choosing
$\lambda_1= \lambda_2= \lambda_3=0$.

\subsection{Feynman rules}
\label{sec:Feynamnrules}
Before moving 
to computation of tree and loop amplitudes,
we set up the corresponding Feynman rules
for supersymmetric Chiral HiSGRAs.

From the Lagrangians  \p{eq:chiralaction} and   \p{eq:chiralaction4}
the propagators are found to be
\be \label{pr-1}
\langle \Theta^{AB}_{\lambda_i}(\pvec_i, p_{\theta,i}),
\Theta^{CD}_{\lambda_j}(\pvec_j, p_{\theta,j}) \rangle
= 
\frac{\delta^{\lambda_i+\lambda_j,-\frac{1}{2}}\delta^4(\pvec_i+\pvec_j)
   \delta(p_{\theta, i}+p_{\theta, j})
   }{\pvec_i^2}\,\Pi_{(G)}^{AB,CD} \quad \text{for} \quad {\cal N}=1, 
\ee
\be \label{pr-4}
\langle \Theta^{AB}_{\lambda_i}(\pvec_i, p_{\theta,i}),
\Theta^{CD}_{\lambda_j}(\pvec_j, p_{\theta,j}) \rangle
=
\frac{\delta^{\lambda_i+\lambda_j,0}
 \delta^4(\pvec_i+\pvec_j)
   \delta^4(p_{\theta, i}+p_{\theta, j})
   }{\pvec_i^2}\,\Pi_{(G)}^{AB,CD} \quad \text{for} \quad {\cal N}=4,
\ee
where the expressions for $\Pi_{(G)}^{AB,CD}$ are  given in \p{Piun}--\p{Piuspn}.

From the interaction terms in \p{eq:chiralaction} and   \p{eq:chiralaction4}
we get for the vertex function
\bea \label{FVer}
{\cal V}(\pvec_i, p_{\theta, i},  \lambda_i) &=&
    \delta^4(\pvec_1+\pvec_2+\pvec_3)
    C_{\lambda_1,\lambda_2,\lambda_3}
   \frac{\PPb^{\lambda_1+\lambda_2+\lambda_3+1}}
    {\beta_1^{\lambda_1 + \frac{1}{2}\epsilon_{\lambda_1}}
    \beta_2^{\lambda_2+ \frac{1}{2}\epsilon_{\lambda_2}}
    \beta_3^{\lambda_3+ \frac{1}{2}\epsilon_{\lambda_3}}} \times \\
    \nonumber
&\times& \int \, \prod_{l=1}^3
dp^{\cal N}_{ \theta,l }\, \,
\delta^{{\cal N}}(p_{\theta, 1} + p_{\theta, 2} + p_{\theta, 3}),    
\eea
where the coupling constants are given by
\p{eq:magicalcoupling}. The value of
 ${\cal N}$ is either $1$
or $4$,   with $\epsilon_{\lambda_i}=0$
for ${\cal N}=4$.
The integration goes over all bosonic momenta, which are not fixed by the momentum
conservation. Besides the vertex function
\p{FVer} is multiplied by wavefunctions of superfields which correspond to external legs.

For the sake of completeness let us note that the Feynman rules given above can be applied
for non-chiral ${\cal D}=4$ ${\cal N}=1$ and ${\cal N}=4$
theories in the light-front gauge,
by including the hermitian conjugate vertex to the one given in \p{FVer},
\be \label{FVer-ach}
{\overline {\cal V}}(\pvec_i, p_{\theta, i},  \lambda_i) =
    \delta^4(\pvec_1+\pvec_2+\pvec_3)  \int \, \prod_{l=1}^3
dp^{\cal N}_{ \theta,l }\, \,
\delta^{{\cal N}}(p_{\theta, 1} + p_{\theta, 2} + p_{\theta, 3}) 
{\cal F}
\, ,    
\ee
with
\bea \label{FVer-ach-1}
&&{\cal F}=  {\overline C}_{\lambda_1, \lambda_2 \lambda_3}
   \frac{\PP^{-\lambda_1-\lambda_2-\lambda_3-\frac{1}{2}}}
    {\beta_1^{-\lambda_1 +\frac{1}{2}\epsilon_{\lambda_1} }
    \beta_2^{-\lambda_2+ \frac{1}{2}\epsilon_{\lambda_2}}
    \beta_3^{-\lambda_3+ \frac{1}{2}\epsilon_{\lambda_3}}} \, 
    \PP_\theta,
\\ \nonumber
&&{\overline C}_{-\lambda_1-\frac{1}{2}, 
-\lambda_2 -\frac{1}{2}, -\lambda_3 -\frac{1}{2}} =
(-)^{\lambda_1+\lambda_2+\lambda_3+\epsilon_{\lambda_2}+1 }
{ C}^\star_{\lambda_1,\lambda_2,
\lambda_3} \quad \text{for} \quad {\cal N}=1
\eea
and
\be \label{FVer-ach-4}
{\cal F}=(-)^{\lambda_1+\lambda_2+\lambda_3}{ C}^\star_{-\lambda_1,-\lambda_2,-\lambda_3}
   \frac{\PP^{-\lambda_1-\lambda_2-\lambda_3+1}}
    {\beta_1^{-\lambda_1 +2 }
    \beta_2^{-\lambda_2+ 2}
    \beta_3^{-\lambda_3+ 2}} \,\, 
    \frac
    {\varepsilon_{\hat i_1, ...,\hat i_4 }}{4!} 
    \PP_\theta^{\hat i_1},...,
    \PP_\theta^{\hat i_4}
    \quad \text{for} \quad {\cal N}=4
\ee
The definition of $\PP_\theta$ is given in \p{pptheta}.
 
The sums of helicities in \p{FVer}
and in \p{FVer-ach-4}
are restricted to be  non-negative
and non-positive integers, respectively. The sum of the helicities in \p{FVer-ach-1} is 
restricted to be a
half-integer less or equal to $-\frac{3}{2}$.

In the present approach we use a perturbation theory in superspaces,
where  the space-time coordinates are extended with Grassmann
momenta
\cite{Dijkgraaf:2002xd}, rather than more commonly used approach, when
the superspace contains Grassmann coordinates \cite{Gates:1983nr,Kovacs:1999fx, Petrov:2001pce}.
One can reformulate the Feynman rules in the Grassmann coordinate space
by performing a Fourier transform.
The choice of the momentum space representation, as well as the choice of the light-front superspace approach
for the perturbation theory
(see \cite{Ananth:2012tf} for computations of  correlation functions
in ${\cal N}=4$ super Yang-Mills), comes naturally, since the cubic interactions
in the vertices depend on  Grassmann momenta. 
  
%%%%%%%%%%%%%%%%%%%%%%%%%%%%%%%

%%%%%%%%%%%%%%%%%%%%%%%%%%%%%%%%%%%%%%%%%%%

%%%%%%%%%%%%%%%%%%%%%%%%%%%%%%%%%%%%%%%%%%%%%%%%%%%%%%%%%%
\section{Tree Amplitudes}
\label{sec:trees}

\subsection{\texorpdfstring{${\cal N}=1$}{N=1}}
%%%%%%%%%%%%%%%%%%%%%%%%%%%%%%%%%%%%%%%%%%%%%%%%%%%%%%%%%%
In this section we consider the tree level diagrams
for $U(N)$ colored supersymmetric HiSGRA.
 The computations are much simpler than in the other versions of HiSGRA, since 
one has to consider only a particular cyclic ordering of external fields.
Consequently, for the four point tree level amplitude one gets two diagrams
 \begin{align*}
  \parbox{2.8cm}{\includegraphics[scale=0.24]{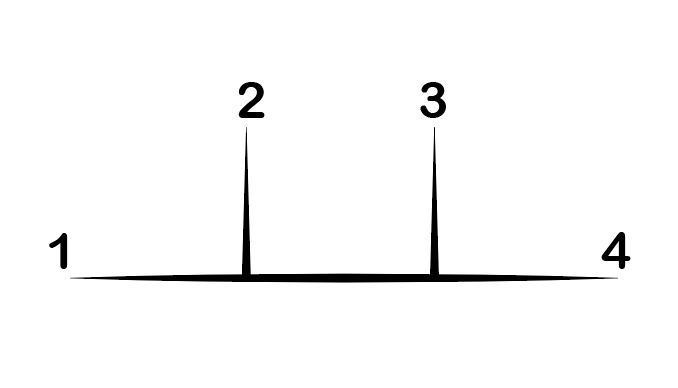}}
  \,\,\,+ \,\,\, \parbox{2.8cm}{\includegraphics[scale=0.24]{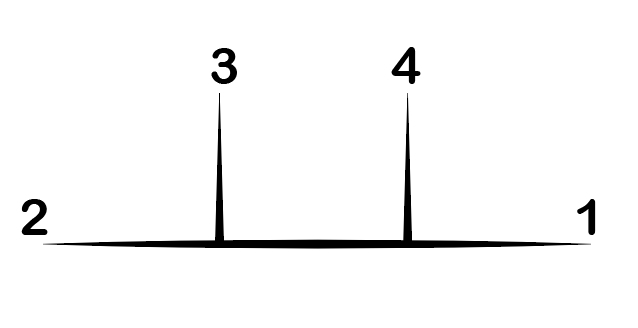}}
\end{align*}
The calculations of tree level amplitudes are nearly identical to the bosonic Chiral HiSGRA \cite{Skvortsov:2018jea}. One modification is due to the fact
that the wave  functions $\Theta_{\lambda_i}$, which are placed on each vertex connected to an external leg, have Grassmann parity  $(-)^{\epsilon_{\lambda_i}}$.
In addition, in the case of ${\cal N}=1$, integration measures $dp_{\theta,i}$ and  the propagator \eqref{pr-1},  are Grassmann-odd. Therefore,  one must pay particular attention to their ordering in order to get the correct sign for each diagram. A general rule for a diagram with $n$ external legs is that the sum of superhelicities 
 for the external legs
\be \label{la}
\Lambda_n \equiv \sum_{i=1}^n \lambda_i
\ee
should be half-integer for even $n$, and integer for odd $n$, else the diagram is zero trivially because of the requirements that superhelicities add up to an integer at each cubic vertex.
 
 Taking into account explicitly
 the sign in the vertices \eqref{eq:magicalcoupling},
 we have for the first diagram in the four point amplitude
\begin{align} \label{eq:sign4a}
A_4(12\veryshortarrow34) = &\int dp_{\theta_1} dp_{\theta_2} dp_{\theta_\omega} \delta( p_{\theta_1}+p_{\theta_2}+p_{\theta_\omega}) (-)^{\epsilon_{\lambda_2}} \Theta_{\lambda_1}\Theta_{\lambda_2} \delta(p_{\theta_\omega}+p_{\theta_{\omega\prime}}) \times \\ \nonumber  \qquad&\times dp_{\theta_{\omega\prime}} dp_{\theta_3} dp_{\theta_4} \delta( p_{\theta_{\omega\prime}}+p_{\theta_3}+p_{\theta_4})(-)^{\epsilon_{\lambda_3}} \Theta_{\lambda_3}\Theta_{\lambda_4} 
\tilde A_4(12\veryshortarrow34) =
\\ \nonumber
=&\,(-)^{1+\epsilon_{\lambda_1}+\epsilon_{\lambda_3}}\int \prod_{i=1}^4 d p_{\theta,i} \, \delta( p_{\theta_1}+p_{\theta_2}+p_{\theta_3}+p_{\theta_4}) \Theta_{\lambda_1}\Theta_{\lambda_2} \Theta_{\lambda_3}\Theta_{\lambda_4} \tilde A_4(12\veryshortarrow34)
\end{align}
where  $\tilde A_4(12\veryshortarrow34)$ is essentially the contribution one gets
for the purely bosonic case
\be
\tilde A_4(12\veryshortarrow34)=
\frac{\PPb_{12}\PPb_{34}
(\PPb_{12}+\PPb_{34})^{\Lambda_4-\frac12}}{4\Gamma(\Lambda_4+\frac12)\prod_{i=1}^{4}\beta_i^{\lambda_i+\frac12\epsilon_{\lambda_i}}(\pvec_1 + \pvec_2)^2}
    \delta^4  (\sum_{i=1}^4 \pvec_i  ) \,.
\ee
For the second diagram we permute the indices cyclically, and then rearrange the integration measure and wave functions to bring them to the same ordering as for the first diagram:
\begin{align}  \label{eq:sign4b}
A_4(23\veryshortarrow41)
= &\,\,(-)^{1+\epsilon_{\lambda_2}+\epsilon_{\lambda_4}}\int dp_{\theta_2} dp_{\theta_3} dp_{\theta_4} dp_{\theta_1} \delta( p_{\theta_2}+p_{\theta_3}+p_{\theta_4}+p_{\theta_1}) \times \\ \nonumber
& \times \Theta_{\lambda_2}\Theta_{\lambda_3} \Theta_{\lambda_4}\Theta_{\lambda_1} {\tilde A}_4(23\veryshortarrow41)
= \\ \nonumber
&= (-)^{\epsilon_{\lambda_2}+\epsilon_{\lambda_4}}\int \prod_{i=1}^4 d p_{\theta,i} \,  \delta( p_{\theta_1}+p_{\theta_2}+p_{\theta_3}+p_{\theta_4}) \Theta_{\lambda_1}\Theta_{\lambda_2} \Theta_{\lambda_3}\Theta_{\lambda_4}
{\tilde A}_4(23\veryshortarrow41),
\end{align}
with
\be
\tilde A_4(23\veryshortarrow41)=
\frac{\PPb_{23}\PPb_{41}
(\PPb_{23}+\PPb_{41})^{\Lambda_4-\frac12}}{4\Gamma(\Lambda_4+\frac12)\prod_{i=1}^{4}\beta_i^{\lambda_i+\frac12\epsilon_{\lambda_i}}(\pvec_2 + \pvec_3)^2}
    \delta^4  (\sum_{i=1}^4 \pvec_i  ) \,.
\ee
Since $\Lambda_4$ is a half-integer, one obtains that $\sum_{i}\epsilon_{\lambda_i}$ is odd. Therefore, the signs of both diagrams are the same and they add up as in the purely bosonic case
\cite{Skvortsov:2018jea}.

Summing both diagrams, and keeping the four-momentum of the first particle off-shell,
\bea
    A_4(1234) &=&
    \frac{(-)^{\epsilon_{\lambda_2}+\epsilon_{\lambda_4}}\alpha_4^{\Lambda_4-\frac12}\beta_3\,\pvec_1^2}{4\Gamma(\Lambda_4+\frac12)\prod_{i=1}^{4}\beta_i^{\lambda_i+\frac12\epsilon_{\lambda_i}-1}\beta_1\PP_{23}\PP_{34}} \times \\ \nonumber
   &\times& \delta^4  (\sum_{i=1}^4 \pvec_i  ) \,
    \int \prod_{j=1}^4  d p_{\theta,j} \,
    \delta (\sum_{k=1}^4 
    p_{\theta,k}  ), \prod_{l=1}^4 \Theta_{\lambda_l}(p_l, p_{\theta_l})
    \eea
    where 
\begin{equation}
\alpha_4=\PPb_{12}+\PPb_{34}.
\end{equation}
After computing the four point tree amplitude, 
one can compute $n$-point diagrams recursively,
using the method of 
\cite{Berends:1987me}.
For example, the five-point function can be computed
by using four- and three-point functions as follows
 \begin{align*}
   \parbox{2.8cm}{\includegraphics[scale=0.3]{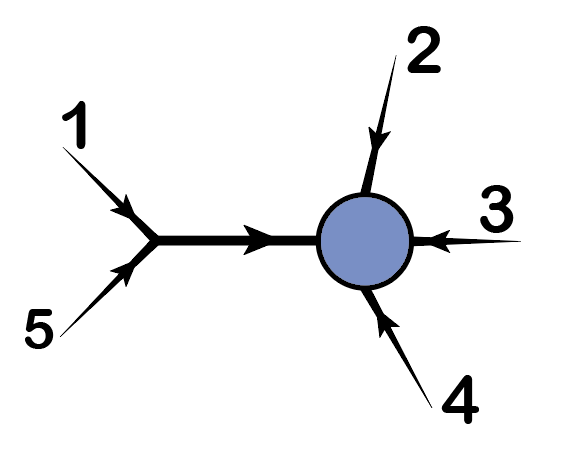}}\,\, + \parbox{3.8cm}{\includegraphics[scale=0.3]{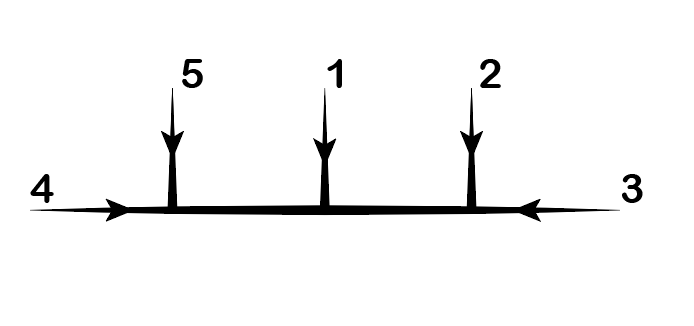}} \,\, + \parbox{3.8cm}{\includegraphics[scale=0.3]{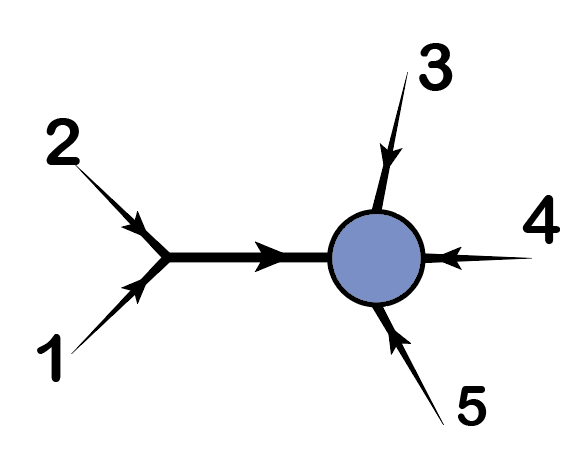}}
\end{align*}
where the external particles
in the four-and three-point functions
which are used as propagators, are kept off-shell.
Note that in the computation of the sign factor 
which comes from the
 four point amplitudes in \eqref{eq:sign4a}-\eqref{eq:sign4b}, one now has to remove the external wave function  $\Theta_{\lambda_1}$ to get the correct result for the amplitude with the first particle being  off-shell. Alternatively, from the same considerations as for the four-point amplitude one can conclude that the $n$-point comb diagram will come with a factor of
$ A_{n} \propto (-)^{\sum_k \epsilon_{\lambda_{2k}}} $
for any cyclic permutation of $123\ldots n$. Then one can sum all the $n$-point diagrams by noting that they have the same structure in the bosonic case.

In this way one can prove  the following expression for
 a tree level $n$--point function
\bea\label{eq:npointrecursive}
    A_n(1...n)&=&\frac{(\prod_{k=1}^{\lfloor n/2 \rfloor} (-)^{\epsilon_{\lambda_{2k}}})(-)^{n}\,\alpha_n^{\Lambda_n-\frac{n-3}2}\beta_3...\beta_{n-1}\,\pvec_1^2}{2^{n-2}\Gamma(\Lambda_n-\frac{n-3}2+1)\prod_{i=1}^{n}\beta_i^{\lambda_i+\frac12\epsilon_{\lambda_i}-1}\beta_1\PP_{23}...\PP_{n-1,n}} \times \\  \nonumber
    &\times& \delta^4  (\sum_{i=1}^n \pvec_i  ) \,
    \int \prod_{j=1}^n d p_{\theta,j} \,\delta(\sum_{k=1}^n
    p_{\theta,k}  )  \prod_{l=1}^n \Theta_{\lambda_l}(p_l, p_{\theta_l}),
    \eea
where
    \begin{equation} \label{alpha-n}
    \alpha_n=\sum_{i<j}^{n-2}\PPb_{ij}+\PPb_{n-1,n}.
\end{equation}
and the four-momentum of the first particle is taken off-shell. When taken on shell, the amplitude vanishes.

\subsection{\texorpdfstring{${\cal N}=4$}{N=4}}
The computations for  tree level $n$-point
amplitudes for the Chiral ${\cal N}=4$ HiSGRA
can be performed in a similar way.
They are however, simpler then for
the case of ${\cal N}=1$ 
since  all superfields, propagators and integration measures are even.
As a result 
the expression for the $n$-point tree level amplitude
obtained  in terms of ${\cal N}=4$ superfields, is almost identical to the case of the bosonic Chiral HiSGRA and reads
\cite{Skvortsov:2018jea,Skvortsov:2020wtf}
\bea\label{eq:npointrecursive4}
    A_n(1...n)&=&\frac{ (-)^{n}\,\alpha_n^{\Lambda_n}\beta_3...\beta_{n-1}\,\pvec_1^2}{2^{n-2}\Gamma(\Lambda_n)\prod_{i=1}^{n}\beta_i^{\lambda_i-1}\beta_1\PP_{23}...\PP_{n-1,n}} \times \\ \nonumber
 &\times&   \delta^4  (\sum_{i=1}^n \pvec_i  ) \, 
    \int \prod_{j=1}^n  d^4 p_{\theta,j}
    \delta^4(\sum_{j=1}^n
    p_{\theta,j}  ) \prod_{l=1}^n \Theta_{\lambda_l}(p_l, p_{\theta_l}),
    \eea
with   $\Lambda_n$ and $\alpha_n$ are defined as in
\p{la} and
\p{alpha-n}, respectively.
%%%%%%%%%%%%%%%%%%%%%%%%%%%%%%%%%%%%%%%%%%%%%%%%%%%%%%%%%%

\section{Loop Amplitudes}
\label{sec:loops}

%%%%%%%%%%%%%%%%%%%%%%%%%%%%%%%%%%%%%%%%%%%%%%%%%%%%%%%%%%
As usually happens in supersymmetric field theories, the tadpole diagrams vanish both for ${\cal N}=1$ and
${\cal N}=4$ Chiral HiSGRAs
due to the property $\delta(0)=0$
 of the Grassmann $\delta$-function, which is present in the propagators \p{pr-1}--\p{pr-4}.

%%%%%%%%%%%%%%%%%%%%%%%%%%%%%
\subsection{Self-energy}
\label{sec:selfenergy1loop}

The simplest one loop diagram corresponds to  the self-energy amplitude  \begin{align*}
   \parbox{3.7cm}{\includegraphics[scale=0.21]{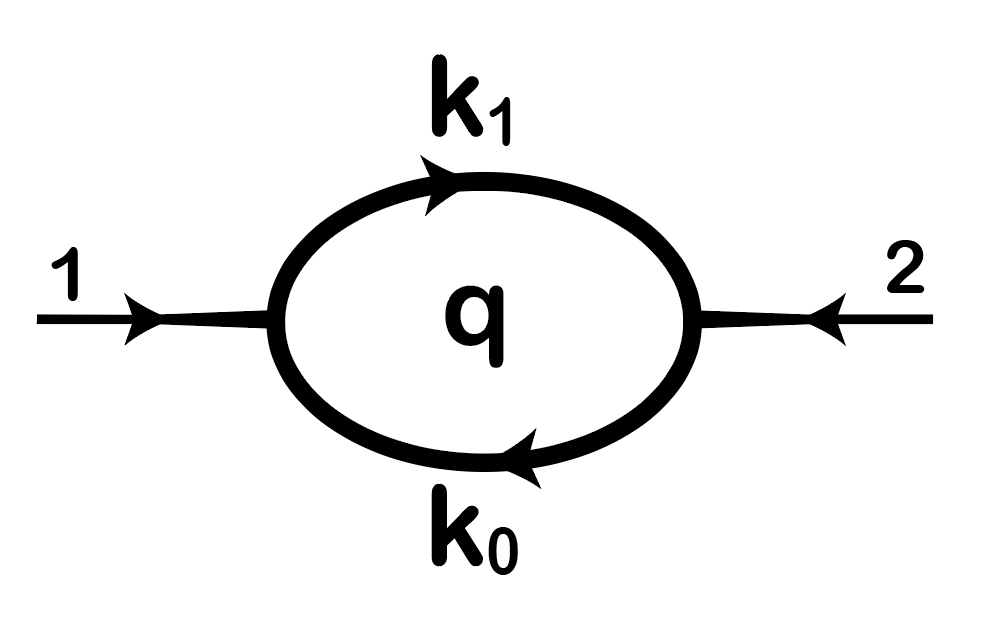}}
\end{align*}
This diagram can be evaluated by introducing of
 dual momenta $\kvec_1,\kvec_0,\qvec$,  related
to the external momentum as
 $\pvec_1=\kvec_1-\kvec_0$. The loop momentum is $\pvec=\qvec-\kvec_0$. 
Using the Feynman rules, given in Section \ref{sec:Feynamnrules},
one can compute for the self-energy diagram
\be \label{self}
\Gamma_{\text{self}} = [\delta(p_{1,\theta}+p_{2,\theta})]^2
\tilde  \Gamma_{\text{self}},
\ee
where, performing a summation over internal helicities $\omega$, one has
\bea\label{eq:selfintegrand}
 \tilde    \Gamma_{\text{self}}&=&N\sum_{\omega} \frac{(l_p)^{\Lambda_2}}
    {\beta_1^{\lambda_1+\frac{1}{2}\epsilon_{\lambda_1}}
    \beta_2^{\lambda_2+\frac{1}{2}\epsilon_{\lambda_2}}\, \Gamma(\Lambda_2)}\int \frac{d^4q}{(2\pi)^4} \frac{\PPb_{q-k_0,p_1}^2\delta_{\Lambda_2,0}}{(\qvec-\kvec_0)^2(\qvec-\kvec_1)^2}= \\ \nonumber
    &=& N\nu_0 ({\bar k}^2_0+ {\bar k}_0{\bar k}_1+ {\bar k}^2_1 )
     \frac{(l_p)^{\Lambda_2}\delta_{\Lambda_2,0}}{\beta_1^{\lambda_1
     +\frac{1}{2}\epsilon_{\lambda_1}
     -1}\beta_2^{\lambda_2+\frac{1}{2}\epsilon_{\lambda_2}-1} \,\Gamma(\Lambda_2)}.
   \eea
The expression $\Gamma_{\text{self}}$ corresponds to 
the self-energy amplitude for the purely bosonic HiSGRA
\cite{Skvortsov:2018jea, Skvortsov:2020wtf}.
It is proportional to 
a finite expression, times
the total number of polarizations $\nu_0$.
In the bosonic case it is an infinite sum
$\nu_0 = 1  +2 \sum_{\lambda=1}^\infty 1$,
where the first $"1"$ stands for the scalar and $"2"$ per each massless
higher spin field. According to the prescription of \cite{Beccaria:2015vaa},
this sum is regularized to zero,
$\nu_0=1 + 2 \zeta_R(0)=0$, by using the Riemann zeta function regularization.

In the ${\cal N}=1$ supersymmetric case, each superfield
carries twice as many degrees of freedom as the bosonic field.
Besides, one has a ''doubling" of the spectrum, discussed in 
subsection \ref{sec:hisgra1}. Similar considerations can be applied for 
${\cal N}=4$ Chiral HiSGRA, where each superfield carries four
times as many degrees of freedom as ${\cal N}=1$ superfield.
In any case, supersymmetry provides further
  cancellations between bosonic and fermionic loops, reflected 
by the presence 
of the square of the Grassmann $\delta$-function
in \p{self}. As long as the number of degrees of freedom can be regularized to a finite value, this makes the amplitude vanish and therefore, supersymmetry is an additional ``source" of finiteness of the Chiral HiSGRA at the one loop level.

\subsection{General Argument for One Loop Amplitudes}
As it has been proven in \cite{Skvortsov:2020gpn},
a general $n$--point one loop amplitude in the bosonic 
Chiral HiSGRA can be obtained by combining
$A_{i}(1,...,i)$ and $A_{n-i}(i+1.,,,n)$ tree amplitudes 
with the self-energy amplitude and taking cyclic permutations.
\begin{align*}\label{eq:recursivegraph}
   \sum_{i=1}^{[n/2]} \parbox{1.8cm}{\includegraphics[scale=0.2]{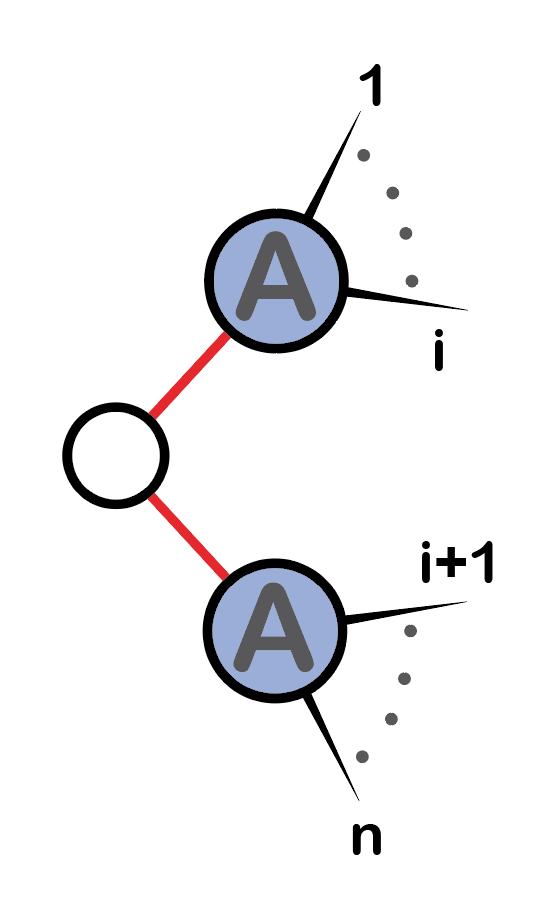}}=\Bigg[\parbox{2.82cm}{\includegraphics[scale=0.27]{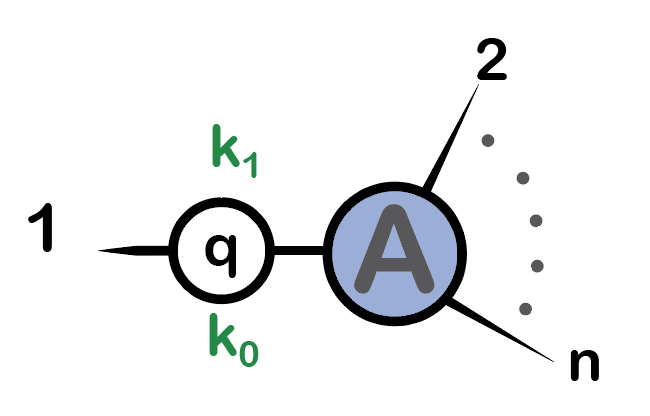}}+\parbox{2.88cm}{\includegraphics[scale=0.27]{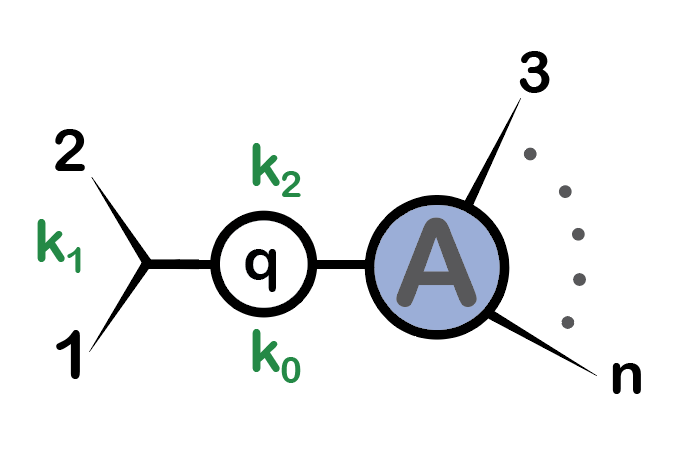}}+\parbox{3.6cm}{\includegraphics[scale=0.265]{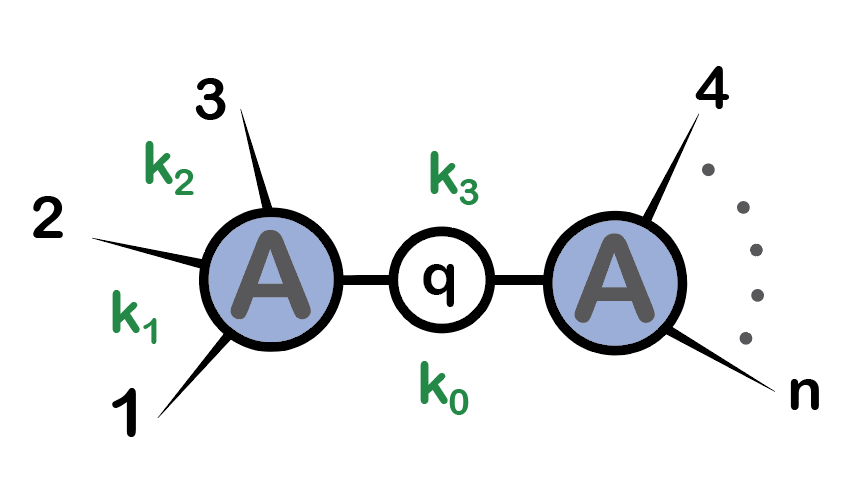}}+...\Bigg]
\end{align*}
This procedure leads in the purely bosonic case
to the general structure of one-loop amplitudes
in the Chiral HiSGRA
\be \label{gen}
\Gamma_{\text{1-loop}}=\Gamma_{\text{1-loop, QCD}}^{++.,,.+} \times D^{HSGR}_{\lambda_1,...\lambda_n} \times \nu_0,
\ee
where $D^{HSGR}_{\lambda_1,...\lambda_n}$ is a kinematical higher spin dressing factor. 
Again, the total amplitude vanishes because the number $\nu_0$
is regularized to zero.
Repeating the same procedure for the supersymmetric case
one can see that supersymmetry provides another cancellation mechanism,
because of the vanishing of individual self-energy
amplitudes that was discussed in the previous subsection.

%%%%%%%%%%%%%%%%%%%%%%%%%%%%%%%%%%%%%%%%%%%
\section{Massive \texorpdfstring{${\cal D}=3$, ${\cal N}=2$}{D = 3, N = 2} Chiral HiSGRA}
\label{D3N2}

As was mentioned in the Introduction, 
 three dimensional massive bosonic Chiral HiSGRA can be obtained
from the four dimensional theory by dimensional reduction \cite{Metsaev:2020gmb}--\cite{Skvortsov:2020pnk}.
In what follows we describe a ${\cal N}=2$ supersymmetric extension of this model,
by making a dimensional reduction 
of 
 the ${\cal D}=4$, ${\cal N}=1$ Chiral HiSGRA,
 considered in the subsection 
 \ref{N1chiralhisgra}.
 We start, by making 
a Fourier transformation with respect to the
$p^2$ component of the four momentum and then making the corresponding
$x^2$
coordinate compact. Then,  we expand ${\cal D}=4$  superfields 
$\Theta_{\lambda}(\pvec, p_\theta)$ as
\be \label{d3SF}
\Theta_{\lambda}(\vec{p}, x^2, p_\theta)=
\sum_{k} \exp{\left ( ikm x^2 \,\varepsilon  \right )} \Theta_{\lambda, k}(\vec{p}, x^2, p_\theta) , \qquad \varepsilon  \equiv \mathrm{sign}(\lambda)
\ee
with $\vec{p}= (\rho, \beta, \gamma)$ being a three-momentum. The mass scale $m$ is determined by the compactification radius.
From the form of the compactification
\p{d3SF}   
it follows, that the masses of the ${\cal D}=3$
superfields have the form $m_k = \varepsilon m k$, for some
integer $k$.
In this way, one obtains ${\cal D}=3$, ${\cal N}=2$
supersymmetry with central charges, as can be seen 
for example from the supersymmetry transformations
 \p{susycompon}.

Inserting the expression
\p{d3SF}
into the action \p{eq:chiralaction}
and integrating over the compact coordinate,
one gets
\bea\label{eq:chiralaction-m}
S&=&-\sum_{\lambda \geq 0, k}
\int d^3 p \, d p_\theta \,\,
(-)^{\epsilon_{\lambda}}
((\vec{p})^2 + m^2 k^2)\,
\mathrm{Tr}[\Theta_{\lambda-\frac{1}{2},k}({\vec{p}, p_\theta})\,  \Theta_{-\lambda,k}({-\vec{p}, -p_\theta})] 
\\ \nonumber
&+&\sum_{\lambda_{1,2,3};  k_{1,2,3}}\int
\prod_{i=1}^3 d^3 p_i  \prod_{j=1}^3 d p_{\theta,j} \,
\delta^3(\vec{p}_1+\vec{p}_2+\vec{p}_3)
    \delta(p_{\theta_1}+p_{\theta_2}+p_{\theta_3})
C(k_i, \lambda_i) \,\,\,  V(k_i, \lambda_i, \vec{p}_i,p_{\theta,i}),
\eea
with the cubic vertex
\be\label{eq:generalvertexD3}
    V= \frac{
    (\PP + \PP_\lambda)^{\lambda_1+\lambda_2+\lambda_3+1}}
    {\beta_1^{\lambda_1 + \frac{1}{2}\epsilon_{\lambda_1}}
    \beta_2^{\lambda_2+ \frac{1}{2}\epsilon_{\lambda_2}}
    \beta_3^{\lambda_3+ \frac{1}{2}\epsilon_{\lambda_3}}}\tr
    [ \Theta_{\lambda_1, k_1}(\vec{p}_1, p_{\theta,1}) \,
 \Theta_{\lambda_2, k_2}(\vec{p}_2, p_{\theta,2}) \,
 \Theta_{\lambda_3, k_3}(\vec{p}_3, p_{\theta,3})],
   \ee
where $\PP$ and $\PP_\lambda$
are given by \p{ppb} and \p{ppl}
respectively,
with the momenta $p$
 having only a real component $\rho$, as it can be seen
 by performing a decomposition of the four dimensional
 complex transverse momenta according to
 $\bar p = p_1- ip_2= \rho - im \varepsilon k$.
   The expression for the coupling constants in \p{eq:generalvertexD3}
\begin{equation}\label{eq:magicalcouplingD3}
    C_{\lambda_1,\lambda_2,\lambda_3}=\frac{(-)^{\epsilon_{\lambda_2}}(l_p)^{\lambda_1+\lambda_2+\lambda_3}
    }{\Gamma(\lambda_1+\lambda_2+\lambda_3+1)} \, \delta_{ k_i \varepsilon_i,0}, \qquad \varepsilon_i \equiv \mathrm{sign}(\lambda_i)
\end{equation}
includes a condition, which implies that the masses of the fields present in the cubic vertex belong to a lattice 
\be
m_1 \varepsilon_1 + m_2 \varepsilon_2 + m_3 \varepsilon_3=0.
\ee
Alternatively,
cubic vertices
for ${\cal D}=3$,  ${\cal N}=2$ supersymmetric massive higher spins  without central charges
can be constructed, using the  method
described in the Appendix  \cite{Metsaev:2021bjh}.

Before proceeding further, let us note,
that the quantum consistency of the higher dimensional
theory is not a priori  preserved by the dimensional reduction, 
and therefore has to be checked separately 
\cite{Fradkin:1982kf}.
To this end,
we set up the corresponding Feynman rules in ${\cal D}=3$
and then  compute  tree and loop amplitudes,
as we did in the previous sections.

The propagator and the vertex functions are
\be \label{pr-1-3d}
\langle \Theta^{AB}_{\lambda_i, k_i}(\vec{p}_i, p_{\theta,i}),
\Theta^{CD}_{\lambda_j,k_j}(\vec{p}_j, p_{\theta,j}) \rangle
= \frac{\delta^{\lambda_i+\lambda_j,\frac{1}{2}}
\delta^{k_i,k_j}
\delta^3(\vec{p}_i+\vec{p}_j)
   \delta(p_{\theta, i}+p_{\theta, j})
   }{\vec{p}_i^2 + m^2 k_i^2}\,\Pi_{(G)}^{AB,CD}, 
\ee
\bea \label{ver-3d}
{\cal V}(\vec{p}_i, p_{\theta,i}\lambda_i, k_i) &=&
   \delta^3(\vec{p}_1+\vec{p}_2+\vec{p}_3)
   C_{\lambda_1,\lambda_2,\lambda_3}
   \frac{(\PP+ \PP_\lambda)^{\lambda_1+\lambda_2+\lambda_3+1}}
    {\beta_1^{\lambda_1 + \frac{1}{2}\epsilon_{\lambda_1}}
    \beta_2^{\lambda_2+ \frac{1}{2}\epsilon_{\lambda_2}}
    \beta_3^{\lambda_3+ \frac{1}{2}\epsilon_{\lambda_3}}} \times  \\ \nonumber
    &\times& \int \, \prod_{l=1}^3
dp_{ \theta,l }\, \,
\delta(p_{\theta, 1} + p_{\theta, 2} + p_{\theta, 3}).    
\eea
Using the Feynman rules one can show, that
tree level $n$--point functions
vanish in complete analogy of the purely bosonic case
\cite{Skvortsov:2020pnk} and with four dimensional
supersymmetric models, considered in Section \ref{sec:trees}.

Finally, let us consider the self-energy diagram. Using 
the Feynman rules \p{pr-1-3d}--\p{ver-3d},
and proceeding in the same way as for the four dimensional
case, one
 obtains the expression
\p{self} with
\bea\label{eq:selfintegrand3d}
    \tilde \Gamma_{\text{self}}&=&  \,
    N\sum_{\omega} \sum_{l} \frac{(l_p)^{\Lambda_2}}
    {\beta_1^{\lambda_1+\frac{1}{2}\epsilon_{\lambda_1}}
    \beta_2^{\lambda_2+\frac{1}{2}\epsilon_{\lambda_2}}\, \Gamma(\Lambda_2)} \times
    \\ \nonumber
    &\times&
    \int \frac{d^3q}{(2\pi)^3} \frac{\PP_{q-k_0,p_1}^2\delta_{\Lambda_2,0}}
    {((\vec{q}-\vec{k}_0)^2+ (l-r_0)^2m^2)
    ((\vec{q}-\vec{k}_1)^2+(l-r_1)^{ 2}m^2)}. 
   \eea
The sum over the integer  $l$ and the appropriate regularization
can be performed using the
approach developed for Kaluza-Klein compactifications
   \cite{Mirabelli:1997aj, Arkani-Hamed:2001jyj, Ghilencea:2001bv}, whereas
  the integral over the non-compact component of the momentum can be treated similarly to the four dimensional case. The main conclusion
  is that supersymmetry makes the entire contribution equal to zero,
  similarly to how it happened for  the analogous diagrams
  in ${\cal D}=4$ Chiral HiSGRAs.

%%%%%%%%%%%%%%%%%%%%%%%%%%%%%%%%%%%%%%%%%%%%%%%%%%%%%%%%%%
%\section{Conclusions}
%\label{sec:conclusions}

%%%%%%%%%%%%%%%%%%%%%%%%%%%%%%%%%%%%%%%%%%%%%%%%%%%%%%%%%%

%%%%%%%%%%%%%%%%%%%%%%%%%%%%%%%%%%%%%%%%%%%%%%%%%%%%%%%%%%
\section*{Acknowledgments}
\label{sec:Aknowledgements}
We are grateful to Yasha Neiman
and Evgeny Skvortsov  
for useful discussions.
The work was supported by the Quantum Gravity Unit
of the Okinawa Institute of Science and Technology Graduate University
(OIST). 
%%%%%%%%%%%%%%%%%%%%%%%%%%%%%%%%%%%%%%%%%%%%%%%%%%%%%%%%%%

%%%%%%%%%%%%%%%%%%%%%%%%%%%%%%%%%%%%%%%%%%%%%%%%%%%%%%%%%%
\begin{appendix}
\renewcommand{\thesection}{\Alph{section}}
\renewcommand{\theequation}{\Alph{section}.\arabic{equation}}
\setcounter{equation}{0}\setcounter{section}{0}
%%%%%%%%%%%%%%%%%%%%%%%%%%%%%%%%%%%%%%%%%%%%%%%%%%%%%%%%%%

%%%%%%%%%%%%%%%%%%%%%%%%%%%%%%%%%%%%%%%%%%%%%%%%%%%%%%%%%%
\section{Cubic Vertices for \texorpdfstring{${\cal N}=1$}{N = 1}}
\label{app:LC}
\setcounter{equation}{0}
%%%%%%%%%%%%%%%%%%%%%%%%%%%%%%%%%%%%%%%%%%%%%%%%%%%%%%%%%%

  In the light-front formulation
one chooses the four dimensional coordinates as
  \be
  x^\pm = \frac{1}{\sqrt 2}(x^3 \pm x^0), \quad
  z = \frac{1}{\sqrt 2}(x^1 + ix^2), \quad
  \bar z = \frac{1}{\sqrt 2}(x^1 - ix^2),
  \ee
  the corresponding components of the four momentum being denoted
  as $\beta, \gamma, p$ and $\bar p$.
 The combinations of momenta that appear in
the cubic interaction vertices and scattering amplitudes have the form
\begin{align}
   \PP_{km}&=p_k\beta_m-p_m\beta_k\,, & \PPb_{km}&=\pb_k\beta_m-\pb_m\beta_k \,. &
   \PP_{km, \theta}&=p_{k, \theta}\beta_m-p_{m, \theta}\beta_k, 
\end{align}
where $k$ and $m$ are numbers of the fields.
For  $n$-point amplitudes only
$n-2$ 
combinations $\PP_{km}$ 
and  $\PPb_{ij}$
are independent due to  the momentum conservation $\sum p_k=\sum \bar p_k= \sum \beta_k=\sum p_{ \theta,k}=0 $. 
For a cubic vertex the only independent combinations are
\bea \label{PPdef}
    \PP&=&
    \frac13\left[ (\beta_1-\beta_2)p_3+(\beta_2-\beta_3)p_1+(\beta_3-\beta_1)p_2\right],  \label{ppb}\\
     \PP_\theta&=&
    \frac13\left[ (\beta_1-\beta_2)p_{\theta, 3}+(\beta_2-\beta_3)p_{\theta, 1}+
    (\beta_3-\beta_1)p_{\theta, 2}\right],   \label{pptheta}
    \eea
and the complex conjugate to \p{ppb}. In ${\cal D}=3$ we use also a combination
\be \label{ppl}
\PP_\lambda=
    \frac{i}{3}\left[ (\beta_1-\beta_2)\varepsilon_3 k_3+(\beta_2-\beta_3)\varepsilon_1 k_1
    +(\beta_3-\beta_1)\varepsilon_2 k_2\right], 
\ee
as well as \p{PPdef} with 
the complex momenta $p_k$ having only a real component $\rho_k$.

Let us move to the construction of the cubic vertices for ${\cal N}=1$ following \cite{Metsaev:2019dqt} and 
refer to \cite{Metsaev:2019aig}
for analogous construction for ${\cal N}=4$.
Recall, that in four dimensions ${\cal N}=1$  super Poincar\'e algebra
without central charges
contains    
generators of Lorentz 
transformations $J^{\mu \nu}$, 
   generators of translations $P^\mu$, and 
  generators of Supersymmetry transformations  $Q^{\alpha}$, and ${\bar Q}^{\dot \alpha}$.
  In order to construct cubic and higher order vertices, one splits the generators  
  of the super Poincar\'e algebra
  into kinematical and dynamical ones
\begin{align}
\text{kinematical}&: && P^{+}, P^z, { P}^{\bar z}, J^{z+}, { J}^{\bar z +}, J^{+-}, J^{z \bar z}, Q^{+},{\bar Q}^{+}, &&\\
\text{dynamical}&: && P^{-}, J^{z-}, { J}^{\bar z-}, Q^{-},{\bar Q}^{-}.   &&
\end{align}
The coordinate $x^+$ is treated as time
and $P^{-}$ as a Hamiltonian.
The kinematical generators at the surface $x^+=0$ are realized
in terms of differential operators as follows
\be \label{KF}
{P}^+=\beta\,, \quad {P}^z=p, \quad P^{\bar z}=\pb,
\quad
{J}^{z+}=-\beta\pfrac{\pb}, \quad  {J}^{\zb+}=-\beta\pfrac{p},
\ee
\be \nonumber
 {J}^{-+}=-\frac{\pl}{\pl \beta} \beta - \frac{1}{2}p_\theta \frac{\partial}{\partial p_{\theta}} + \frac{1}{2} \epsilon_\lambda,
\quad {J}^{z\zb}= p\pl_p-\bar p\frac{\partial}{\partial \bar p} +\lambda -\frac{1}{2}p_\theta \frac{\partial}{\partial p_{\theta}} 
\ee
\be \nonumber
Q^+ = (-)^{\epsilon_\lambda} \beta \frac{\partial}{\partial p_\theta}, \quad \bar Q^{+}= (-)^{\epsilon_\lambda} p_\theta
\ee
  For the dynamical operators at $x^+=0$ one has
\bea \label{DF}
    H&=&-\frac{p\pb}{\beta}\,,  \\ \nonumber
        {J}^{z-}&=& -\pfrac{\pb} \frac{ p\pb}{\beta} +p \pfrac{\beta}  \nonumber
        - \left (\lambda - \frac{1}{2} p_\theta \frac{\partial}{\partial p_\theta} \right ) \frac{p}{\beta} 
        + \left( \frac{1}{2} p_\theta \frac{\partial}{\partial p_\theta }-\frac{1}{2}\epsilon_\lambda \right) \frac{ p}{\beta},
        \\ \nonumber
         {J}^{\zb-}&=&- \pfrac{p} \frac{ p\pb}{\beta} +\pb \pfrac{\beta}  
          + \left (\lambda - \frac{1}{2} p_\theta \frac{\partial}{\partial p_\theta } \right ) \frac{\bar p}{\beta} 
        + \left( \frac{1}{2} p_\theta
        \frac{\partial}{\partial p_\theta }-\frac{1}{2}\epsilon_\lambda \right) \frac{ p}{\beta}, \\    \nonumber
    Q^{-} &=&(-)^{\epsilon_\lambda}\frac{p}{\beta} p_\theta, \qquad
    \bar Q^- =(-)^{\epsilon_\lambda} \bar p \frac{\partial}{\partial p_\theta}. \nonumber
    \eea
    From the explicit realization of the 
    generators 
    \p{KF}-\p{DF} and the explicit form of the higher spin 
    superfields \p{SFN1-1}, \p{SFN1-2}, \p{superfieldN4} one can find 
    transformation rules for the component fields.
    For example for supersymmetry transformations we have
    \be \label{susycompon}
    \delta \phi_{s}= 
    (\bar \epsilon^{-} + \frac{ p}{\beta} \,
     \epsilon^{+} )\phi_{s-\frac{1}{2}}, \qquad
    \delta \phi_{s-\frac{1}{2}}= 
    -(\bar \epsilon^{-} \beta +  \epsilon^{+} 
    { \bar p})\phi_{s}
    \ee
    as well as similar expressions of the component fields
    in \p{SFN1-2}. 
    
    For the subsequent calculations it is simpler to
    perform a partial Fourier transform with respect to $\gamma$
    and consider the fields on the surface $x^+=0$.
    To quadratic order in superfields, the Poincar\'e algebra is realised by the expressions
    \be
    G_{2} = \sum_{\lambda= -\infty}^\infty
    \int \beta \, d^3p \ d p_\theta 
    (-)^{\epsilon_{\lambda}}
\mathrm{Tr}[\Theta_{\lambda-\frac{1}{2}}({\pvec, p_\theta})\, G \, \Theta_{-\lambda}({-\pvec, -p_\theta})], 
    \ee
    where $d^3p= d\beta \, dp \,  d \bar p$ and $G$  collectively denotes differential operators given  in \p{KF}-\p{DF}.
    
In the chiral theory one keeps the operators
${\bar Q}^{-}$ and $J^{\zb-}$
quadratic in the fields  also at the interaction level,
and modifies the other dynamical generators as
\be \label{cubiclc-1}
    H_3=H_2+\int d \Gamma_{[3]} \, 
     \Theta^{\lambda_1 \lambda_2 \lambda_3}_{q_1 q_2 q_3} \,
    h_{\lambda_1 \lambda_2 \lambda_3}^{q_1 q_2 q_3}\,,
\ee
\be
     Q_3^{-}=Q_2^{-}
    +\int d \Gamma_{[3]} \,
     \Theta^{\lambda_1 \lambda_2 \lambda_3}_{q_1 q_2 q_3}\,
    q_{\lambda_1 \lambda_2 \lambda_3}^{q_1 q_2 q_3,
    }  \label{cubiclc-3}\,,
    \ee
    \bea
    J^{z-}_3&=&J^{z-}_2+ 
    \int d \Gamma_{[3]} \times  
    \\ \nonumber
    &\times&\left[ 
     \Theta^{\lambda_1 \lambda_2 \lambda_3}_{q_1 q_2 q_3} \,\,
    j_{\lambda_1 \lambda_2 \lambda_3}^{q_1 q_2 q_3}-
    \frac{1}{3}
    \left(\sum_{k=1}^3 \frac{\partial 
     \Theta^{\lambda_1 \lambda_2 \lambda_3}_{q_1 q_2 q_3}
    }{\partial \bar{q}_k}\right)h_{\lambda_1 \lambda_2\lambda_3}^{q_1 q_2 q_3}-
     \frac{1}{3}
     \left(\sum_{k=1}^3 
     \frac{\partial
      \Theta^{\lambda_1 \lambda_2 \lambda_3}_{q_1 q_2 q_3}
     }{\partial {q}_{\theta, k}}\right) q_{\lambda_1 \lambda_2\lambda_3}^{q_1 q_2 q_3}
    \right]\,. 
    \eea
Here $ \Theta^{\lambda_1 \lambda_2 \lambda_3}_{q_1 q_2 q_3} \equiv
 \Theta_{\lambda_1}(\qvec_1, q_{\theta,1}) 
 \Theta_{\lambda_2}(\qvec_2, q_{\theta,2})
 \Theta_{\lambda_3}(\qvec_3, q_{\theta,3})
 $ and
\be \label{measure}
d \Gamma_{[3]}= d \Gamma_{[3,q]} \cdot d \Gamma_{[3,\theta]} =
(2 \pi)^3  \prod_{k=1}^3 \frac{d^3 q_k}{(2 \pi)^\frac{3}{2}} 
\delta^3\left(\sum_{i=1}^3 q_i\right) \cdot
\prod_{l=1}^3 dq^{\cal N}_{ \theta,l }\,\delta^{{\cal N}}\left(\sum_{j=1}^3 q_{\theta, j}\right)
\ee
is an integration measure,
written for a generic
${\cal N}$.
The  vertices
are determined from the requirement
of  preservation of the  super Poincar\'e
algebra at the cubic level in the superfields \cite{Metsaev:2019dqt},   \cite{Metsaev:2019aig}. Their explicit form is found to be
\be \label{lcs1}
h_{\lambda_1 \lambda_2 \lambda_3}^{q_1 q_2 q_3} = C^{\lambda_1 \lambda_2 \lambda_3} 
(\PPb)^{\lambda_1+\lambda_2+\lambda_3+1} 
\prod_{i=1}^3
\beta_i^{-\lambda_i - \frac{1}{2}\epsilon_{\lambda_i}}\,,
\ee
\be \label{lcs2}
q_{\lambda_1 \lambda_2 \lambda_3}^{q_1 q_2 q_3,
} =- C^{\lambda_1 \lambda_2 \lambda_3} 
(\PPb)^{\lambda_1+\lambda_2+\lambda_3} \PP_\theta
\prod_{i=1}^3
\beta_i^{-\lambda_i - \frac{1}{2}\epsilon_{\lambda_i}}\,,
\ee
\be \label{lcs3}
j_{\lambda_1 \lambda_2 \lambda_3}^{q_1 q_2 q_3} =2 C^{\lambda_1 \lambda_2 \lambda_3} 
(\PPb)^{\lambda_1+\lambda_2+\lambda_3}  \, \chi
\prod_{i=1}^3
\beta_i^{-\lambda_i - \frac{1}{2}\epsilon_{\lambda_i}}\,,
\ee
where
\be
    \chi=\beta_1(\lambda_2-\lambda_3)+\beta_2(\lambda_3-\lambda_1)+\beta_3(\lambda_1-\lambda_2)\,.  \ee
As usual, the explicit form of the coupling constants
$C^{\lambda_1 \lambda_2 \lambda_3} $
is not determined at the level of cubic interactions.
The restriction on the coupling constants comes from the further requirement,
 that the (super)Poincar\'e algebra is preserved at all orders 
in (super)fields, without 
adding of quartic and higher order vertices to the dynamical generators. 
In other words, one  has 
to find an expression for coupling constants,
that keeps the equations
\begin{equation}\label{eq:Dconstraint}
    [Q^{-}_3,P^{-}_3]=0, \quad
[J^{z,-}_3,P^{-}_3]=0
\end{equation}
intact also at the quartic level.
Let us consider the first equation. 
Using the expressions 
\p{lcs1} and \p{lcs2}, one can see that
\bea \label{commut}
[Q^{-}_3,P^{-}_3] &\sim& \PPb\, \sum_{\lambda_1 \lambda_2  \lambda_3} \sum_{\tau_1 \tau_2  \tau_3}
C^{\lambda_1 \lambda_2  \lambda_3} C^{\tau_1 \tau_2  \tau_3} \times \\ \nonumber
&\times& \left [\prod_{i=1}^3
\beta_i^{-\lambda_i - \frac{1}{2}\epsilon_{\lambda_i}} \Theta^{\lambda_1 \lambda_2  \lambda_3} \PP^{\lambda_1 + \lambda_2+ \lambda_3}, \prod_{i=1}^3
\beta_i^{-\tau_i - \frac{1}{2}\epsilon_{\tau_i}}
\Theta^{\tau_1 \tau_2  \tau_3} \PP^{\tau_1 + \tau_2 + \tau_3} \right ]
\eea
which is zero due to the antisymmetry of the Poisson bracket.

Now let us consider the second equation in \p{eq:Dconstraint}.
In the same way as for the bosonic Chiral HiSGRA \cite{Skvortsov:2020wtf},
one can show that
the sum of the first two terms
in the nonlinear part of $J_3^{z-}$  gives zero Poisson bracket
with $P^{-}_3$, provided  the coupling constants
 have the form \p{eq:magicalcoupling}.
After integrating  by parts
in the third term
 of the nonlinear part 
of $J_3^{z-}$, one can see, that  
its Poisson  bracket with
 $P^{-}_3$ is zero  by the same argument as in \p{commut}.

%%%%%%%%%%%%%%%%%%%%%%%%%%%%%%%%%%%%%%%%%%%%%%%%%%%%%%%%%%

%%%%%%%%%%%%%%%%%%%%%%%%%%%%%%%%%%%%%%%%%%%

%%%%%%%%%%%%%%%%%%%%%%%%%%%%%%%%%%%%%%%%%%%%%%%%%%%%%%%%%%
\end{appendix}
%%%%%%%%%%%%%%%%%%%%%%%%%%%%%%%%%%%%%%%%%%%%%%%%%%%%%%%%%%

\setstretch{1.0}
\bibliographystyle{JHEP-2}
\bibliography{megaHS.bib}

\end{document}